\documentstyle[11pt,paspconf,epsf]{article}
\def\spose#1{\hbox to 0pt{#1\hss}}
\def\simlt{\mathrel{\spose{\lower 3pt\hbox{$\mathchar"218$}}
     \raise 2.0pt\hbox{$\mathchar"13C$}}}
\def\simgt{\mathrel{\spose{\lower 3pt\hbox{$\mathchar"218$}}
     \raise 2.0pt\hbox{$\mathchar"13E$}}}

\def\edcomment#1{\iffalse\marginpar{\raggedright\sl#1\/}\else\relax\fi}
\reversemarginpar

\begin{document}
\title{Star Formation Rate at $z=0.2$ derived from H$\alpha$
luminosities: constraint on the reddening} \author{Laurence Tresse and
Steve J. Maddox} \affil{Institute of Astronomy, Madingley Road,
Cambridge CB3 0HA, UK}

\begin{abstract}
We discuss the relative merits of using UV and H$\alpha$ as star
formation indicators from galaxy surveys. In particular, comparing UV
and H$\alpha$ in the CFRS gives a limit of a factor 2.5 for the
UV(2800 \AA) extinction from dust, using the conversion factors of
Madau et al. 1998 (Salpeter IMF, 0.1$-$125 M$\odot$). Our strong
correlation between B and H$\alpha$ argues for a universal IMF slope.
The H$\alpha$ LF at $z=0.2$ shows a faint end slope of $\alpha=-1.35$,
which is consistent with fading of short bursts of star formation. We
also discuss the contribution of AGN to UV and H$\alpha$ luminosities. 
\end{abstract}

\section{Luminosity density and star formation rate}

The cosmological evolution of the star formation rate (SFR) can be
inferred from the luminosity density of the galaxy population seen at
different redshifts (Madau et al. 1998).  Indeed the luminosity
density history traces the {\it global} evolution of the galaxy
population, or in other words, of the whole content of galaxies (gas,
stars, dust, active galactic nucleus (AGN), supernovae remnants,
etc.), and not simply of the stellar population.  Recovering the {\it
genuine shape} of the SFR history is essential to differentiate
between scenarios of galaxy evolution and formation.  Extensive
studies of nearby galaxies have shown that the UV, FIR, radio
continuum and H$\alpha$ line fluxes are closely connected (Buat 1992),
even though these radiations are produced by different physical
processes. Thus they originate from a common photoionizing spectrum.
Retrieving the {\it absolute} photoionizing flux depends crucially on
the understanding of these processes.
 
\section{Dust extinction of UV and H$\alpha$ emission} 

Assuming there is no AGN, the galactic UV continuum radiation directly
traces the UV stellar spectra. From the Lyman break (912 \AA) to
$\sim$ 3500 \AA, the emissivity of the starburst galaxy population
becomes less dominated by the relatively massive, young stars (O, B
types, T$\ge$10 000 K, M$\ge 10$ M$_\odot$, t$_{MS}\sim$10$^7$ yr),
while it becomes more dominated by the intermediate mass stars (A
type, M$=2-5$ M$_\odot$, t$_{MS}\sim 10^8$ yr).  The emissivity of the
more quiescent galaxy population (i.e. ellipticals) may be dominated
by massive, old stars (t$\gg$ 10$^9$ yr), but other possibilities, of
which some are non stellar, have been proposed to account for the UV
rise seen in these galaxies. Assuming there is no AGN, the
H$\alpha$($\lambda 6563$ \AA) fluxes come from the hydrogen gas
surrounding very massive, short-lived stars (OB type, T$\sim$ 60 000
K, M$\gg 10$ M$_\odot$, t$_{MS}\sim 10^6$ yr). The gas is directly
excited by far-UV stellar photons shorter than 912 \AA, and its
recombination produces spectral emission lines.  Of the Balmer lines,
H$\alpha$ is the most proportional to the far-UV stellar spectra,
because it is the easiest to ionize, and is barely, if at all,
affected by the stellar absorption due to old stars.  Moreover, it
does not depend on the metal fraction present in the gas, or on the
hardness of the ionizing stellar spectrum (which is the case for metal
lines e.g. [O~II]$\lambda$3727).

The global contribution of these stellar populations depends mainly on
the shape of the initial mass function (IMF) and its mass cut-off, the
average age and metallicity, and the luminosity function (LF) of the
galaxy population considered. Uncertainties arise for ellipticals, for
which the UV source is not fully understood. Also H$\alpha$ fluxes for
galaxies dominated by an old stellar population are likely to be
affected by stellar absorption.  However, in the studies of global
evolution of the galaxy population, the starburst galaxies 
dominate the UV radiation, and so the remaining galaxies are
unlikely to affect the general results.

Recovering the {\it absolute} SFR from UV continuum and H$\alpha$ line
flux measurements depends on the dust extinction, which is subject to
debate.  Indeed massive-star formation occurs in dusty molecular
clouds; most of the UV radiation is reprocessed by dust and emitted in
the far infra-red. H$\alpha$, and UV fluxes arise from relatively late
stages of star formation ($\simgt 10^6$ yr), when the star-forming
region becomes less opaque.  In the optical, different extinction laws
behave similarly, thus retrieving the original H$\alpha$ fluxes is not
a major problem. In the UV, which is more dust affected, they can
differ by a large factor. Thus, it leads to large uncertainties in the
absolute UV radiation.  But, long-lived stars as traced by the
near-UV, sit in less obscured regions than short-lived stars as traced
by H$\alpha$.  This is corrobated by the fact that extinction
measurements from the UV stellar continuum are found to be no larger
(as expected by any extinction laws) than from the optical
H$\alpha$/H$\beta$ decrement.  Consequently, the observed near-UV is
dominated by intermediate mass stars not only because they are more
numerous and longer-lived than massive stars, but also because they are
much less dust obscured.  In summary, H$\alpha$ and near-UV radiations
emerge in average from different environments within a galaxy, because
each is dominated by different stellar populations and time scales
(see e.g. Calzetti et al. 1994). Nevertheless, they are tightly
correlated (Buat et al. 1987) - asserting the universality of the upper
IMF limit.

The relationships between quantities have been extensively studied
in the nearby universe. However, we need to ascertain these physical
parameters with cosmic time.  Clearly the emissivity of galaxies
evolves with redshift, but is the stellar content entirely
responsible?  Or does the dust extinction vary?  Does the AGN
contribution increase?  We need to answer these questions before
getting the correct shape of the SFR. This is essential not only for
including these processes in evolutionary models, but also for
quantifying how much it affects the galaxy selection in deep surveys.
Correlating H$\alpha$ and UV at any redshift is a powerful way to
tackle the dust uncertainties.

\section{SFR from H$\alpha$ at $z=0.2$: dust upper limit}

H$\alpha$ is seen in the optical up to $z\sim0.3-0.4$.  Prospective
instruments in the near infra-red will soon enable systematic studies
of H$\alpha$ and UV radiation within the {\it same} galaxy sample at
$z\gg 0.3$. This will avoid discrepancies inherent to comparing
different surveys.  With the I-selected CFRS galaxies up to $z=0.3$,
we measured the dust-corrected H$\alpha$ luminosities, and obtained a
H$\alpha$ luminosity function (see Tresse \& Maddox 1998 for further
details). We used the factors from Madau et al. 1998 (Salpeter IMF,
0.1$-$125 M$_{\odot}$) to convert our H$\alpha$ luminosity density at
$\langle z \rangle = 0.2$ of 10$^{39.44\pm0.04}$ erg/s/Mpc$^3$ into a
$\log SFR/M_{\odot} yr^{-1} = -1.71\pm 0.04$. At $z<0.3$, the B-band
barely samples the rest-frame UV. Thus we could not compare directly
our H$\alpha$ determination with a UV one without extrapolation, which
is likely to introduce untestable uncertainties. However the B-band of
CFRS galaxies at $0.4<z<1.3$ allowed the determination of the
rest-frame near-UV(2800 \AA).  If we consider the evolution of the
CFRS luminosity density at 2800 \AA\ of $(1+z)^{3.9\pm0.75}$
(h$_{50}=1$, $\Omega=1$, $\lambda=0$), as defined in Lilly et
al. 1996, then the $\log SFR/M_{\odot} yr^{-1}$ from UV(2800 \AA) at
$z=0.2$ should be $-2.13\pm0.14$. Assuming that all parameters are
correct (IMF, stellar populations, no AGN, case B recombination, etc.)
and that the slope $\alpha$ for the star-forming LFs is constant, then
the UV(2800\AA) fluxes are low by a factor $\sim$2.5 (or 1 mag) at
$z=0.2$.  We interpret this as the dust correction required for these
UV measurements.  We note that our {\it average} dust correction from
the H$\alpha$/H$\beta$ decrement is A$_V =$ 1 mag, which is of the
same order as we derived for the UV(2800 \AA) magnitudes at $z\simeq
0.2$. This result is similar to local observations as discussed in
Section 2.

\section{H$\alpha$ luminosities, B-band emissivity, and colors}  

We find that B-band luminosities are tightly correlated to H$\alpha$
luminosities. This is as the correlation between UV and H$\alpha$; the
B band is still dominated by young stars (type A).
This surely reinforces the hypothesis of a universal IMF: for a
certain amount of massive OB stars formed, there is always the same
among of intermediate mass, type A, stars.  We find $M(B_{AB}) = 46.7
-1.6\log L(H\alpha)$, with the luminosities L(H$\alpha$) being dust
corrected, but not the B magnitudes.  If the latter are corrected,
according our results, a dust correction of 0.6 mag in average
(Seaton's law assumed) should be taken.  This relation implies that
surveys at high $z$, which select preferentially bright, star-forming
galaxies, are also likely to pick up only the strongest H$\alpha$
emitters.  We do not find a correlation between H$\alpha$
luminosities, and the rest-frame colors ($B-R$). This endorses the
idea that H$\alpha$ production 
depends mainly on the ``instantaneous'' star formation, or in
other words, on the time scale since the last burst.

\section{The H$\alpha$ luminosity function} 

The H$\alpha$ LF is related to the number of ionizing photons emitted
by massive stars. Since the latter have a short life ($<10^6$ yr),
their number traces the ``instantaneous'' SFR.  The SFR deduced only
from H$\alpha$ luminosities is very dependent on the assumed IMF,
since only the massive stars are traced. The slope of the H$\alpha$ LF
depends more on the time scale since the last star bursts, rather than
on the SFR. A fading process of H$\alpha$ photons produces a non flat
slope. The shallower the IMF, the slower is the fading process, and
steeper is the slope (Hogg \& Phinney 1997).  Because of the tight
relation between B or UV magnitudes, and H$\alpha$ luminosities, the
slope in B- and UV-band LFs must be correlated to the slope of the
H$\alpha$ LF. We note that for the luminosity density history (Madau
et al. 1998, Lilly et al. 1996), a constant slope for star-forming
galaxies has been assumed ($\alpha=-1.3$) at all redshifts, i.e.  a
constant fraction of dwarf galaxies has been considered.  If the
merging rate, and/or the SFR are not constant through cosmic time as
predicted, this assumption has to be revised.  The SFR history is
deduced from the luminosity densities, $\phi^{*}\ L^{*}\ \Gamma(\alpha
+ 2)$; a $(1+z)^x$ evolution in $L^{*}$, or in $\phi^{*}$, is
approximatively equivalent to a slope evolution of $\alpha =
\Gamma^{-1}\left((1+z)^{x} \Gamma(\alpha_0 +2)\right) -2 \simeq
\alpha_{0}+2 / (1+z)^x -2$.

The best fit for our H$\alpha$ LF is $\alpha = -1.35\pm0.06 $,
$\phi^{\ast} = 10^{-2.83\pm0.09}$ Mpc$^{-3}$, and $L^{\ast} =
10^{42.13\pm0.13}$ erg s$^{-1}$.  We point out that the tight relation
found between B or UV magnitudes, and H$\alpha$ luminosities validates
our use of a magnitude-selected survey, like the CFRS, to measure the
H$\alpha$ LF.  We did not exclude the AGN galaxies from our sample to
be able to compare our result to the CFRS rest-frame UV data
(contrarily to Gallego et al. 1995). There are many problems arising
if AGN are excluded.  In principle, they should be excluded, but in
practice, it requires an {\it objective} technique to do so at all
redshifts, in particular for the narrow-line AGN.  Moreover, observing
the whole galaxy content (or part of) as done in deep redshift
surveys, blurs the objective line-ratio classification because
different stellar and dust contents within individual galaxies are
sampled, which leads to overlap between AGN and starbursts (see Tresse
et al. 1996).  In addition, the H$\alpha$ fluxes observed from an AGN
galaxy is produced both by the AGN and the H~II regions.  The
contribution of each ionizing source to the H$\alpha$ flux is unknown.
Thus removing AGN galaxies from a sample underestimates the total
H$\alpha$ flux produced by stars.  Conversely including them provides
an upper-limit of H$\alpha$ luminosities produced by stars.  Also, if
merging processes trigger nuclear activity, then excluding AGN may
lead to the exclusion of this class of galaxies.

\end{document}